\documentclass[aps,prc,twocolumn,showpacs,superscriptaddress,amsmath,amssymb]{revtex4}

\usepackage{graphicx}
\usepackage{dcolumn}
\usepackage{bm}
\begin{document}


\title{Measurement of the Gamow-Teller Strength Distribution in $^{58}$Co via the $^{58}$Ni(t,$^3$He) reaction at 115 MeV/nucleon}

\author{A.L. Cole}
\altaffiliation[Present address: ]{Department of Physics, University of Michigan, Ann Arbor, MI 48109-1040, USA}
\affiliation{National Superconducting Cyclotron Laboratory, Michigan State University, East Lansing, MI 48824-1321, USA}
\affiliation{Joint Institute for Nuclear Astrophysics, Michigan State University, East Lansing, MI 48824, USA}

\author{H. Akimune}
\affiliation{Department of Physics, Konan University, 8-9-1 Okamoto Higashinda, Kobe, Hyogo, 658-8501, Japan}

\author{Sam M. Austin}
\affiliation{National Superconducting Cyclotron Laboratory, Michigan State University, East Lansing, MI 48824-1321, USA}
\affiliation{Joint Institute for Nuclear Astrophysics, Michigan State University, East Lansing, MI 48824, USA}
\affiliation{Department of Physics and Astronomy, Michigan State University, East Lansing, MI 48824, USA}

\author{D. Bazin}
\affiliation{National Superconducting Cyclotron Laboratory, Michigan State University, East Lansing, MI 48824-1321, USA}
\author{A.M. van den Berg}
\affiliation{Kernfysisch Versneller Instituut, University of Groningen, Zernikelaan 25, 9747 AA Groningen, The Netherlands}
\author{G.P.A. Berg}
\affiliation{Department of Physics, University of Notre Dame, IN 46556-5670, USA}
\affiliation{Joint Institute for Nuclear Astrophysics, University of Notre Dame, IN 46556-5670, USA}
\author{J. Brown}
\affiliation{Department of Physics, Wabash College, Crawfordsville, IN 47933, USA}

\author{I. Daito}
\affiliation{Kansai Photon Science Institute, Japan Atomic Research Institute, Kizu Kyoto 619-0215, Japan}

\author{Y. Fujita}
\affiliation{Department of Physics, Osaka University, Toyonaka, Osaka 560-0043, Japan}

\author{M. Fujiwara}
\affiliation{Kansai Photon Science Institute, Japan Atomic Research Institute, Kizu Kyoto 619-0215, Japan}
\affiliation{Research Center for Nuclear Physics, Osaka University, Ibaraki, Osaka 567-0047, Japan}

\author{S. Gupta}
\affiliation{National Superconducting Cyclotron Laboratory, Michigan State University, East Lansing, MI 48824-1321, USA}
\affiliation{Joint Institute for Nuclear Astrophysics, Michigan State University, East Lansing, MI 48824, USA}
\author{K. Hara}
\affiliation{Research Center for Nuclear Physics, Osaka University, Ibaraki, Osaka 567-0047, Japan}

\author{M.N. Harakeh}
\affiliation{Kernfysisch Versneller Instituut, Zernikelaan 25 9747 AA Groningen, The Netherlands}



\author{J. J\"{a}necke}
\affiliation{Department of Physics, University of Michigan, Ann Arbor, MI 48109-1040, USA}

\author{T. Kawabata}
\affiliation{Center for Nuclear Study, University of Tokyo, Bunkyo, Tokyo 113-0033, Japan}

\author{T. Nakamura}
\affiliation{Tokyo Institute of Technology, 2-12-1 O-Okayama, Tokyo 152-8550, Japan}
\author{D.A. Roberts}
\affiliation{Department of Physics, University of Michigan, Ann Arbor, MI 48109-1040, USA}

\author{B.M. Sherrill}
\affiliation{National Superconducting Cyclotron Laboratory, Michigan State University, East Lansing, MI 48824-1321, USA}
\affiliation{Joint Institute for Nuclear Astrophysics, Michigan State University, East Lansing, MI 48824, USA}
\affiliation{Department of Physics and Astronomy, Michigan State University, East Lansing, MI 48824, USA}
\author{M. Steiner}
\affiliation{National Superconducting Cyclotron Laboratory, Michigan State University, East Lansing, MI 48824-1321, USA}

\author{H. Ueno}
\affiliation{Applied Nuclear Physics Laboratory, RIKEN, Wako, Saitama 351-0198, Japan}
\author{R.G.T. Zegers}
\email{zegers@nscl.msu.edu}
\affiliation{National Superconducting Cyclotron Laboratory, Michigan State University, East Lansing, MI 48824-1321, USA}
\affiliation{Joint Institute for Nuclear Astrophysics, Michigan State University, East Lansing, MI 48824, USA}
\affiliation{Department of Physics and Astronomy, Michigan State University, East Lansing, MI 48824, USA}

\date{\today}

\begin{abstract}

Electron capture and $\beta$ decay play important roles in the evolution of 
pre-supernovae stars and their eventual core collapse.  These rates are normally predicted through shell-model calculations. Experimentally determined strength distributions from charge-exchange reactions are needed to test modern shell-model calculations.   
We report on the measurement of the Gamow-Teller strength 
distribution in $^{58}$Co from the $^{58}$Ni(t,$^3$He) reaction with a secondary triton beam of 
an intensity of ${\sim}10^6$ pps at 115 MeV/nucleon and a resolution of 
${\sim}250$ keV. Previous measurements with the $^{58}$Ni(n,p) and the $^{58}$Ni(d,$^2$He) reactions were inconsistent with each other. Our results support the latter. We also compare the results to predictions of large-scale shell model calculations using the KB3G and GXPF1 interactions and investigate the impact of differences between the various experiments and theories in terms of the weak rates in the stellar environment. Finally, the systematic uncertainties in the normalization of the strength distribution  extracted from $^{58}$Ni($^{3}$He,t) are described and turn out to be non-negligible due to large interferences between the $\Delta L=0$, $\Delta S=1$ Gamow-Teller amplitude and the $\Delta\text{L}=2$, $\Delta\text{S}=1$ amplitude.

\end{abstract}


\maketitle

\section{Introduction}
\label{sec:intro}

Weak interactions play an important role in a variety of astrophysical phenomena.  In particular the 
evolution of massive stars during their pre-supernova stage (type-II or core-collapse supernovae) is strongly affected by electron capture and $\beta$ decay rates. Electron-capture leads to deleptonization of the stellar environment. The dynamics of the collapse process is modified and the isotopic composition of the star is changed \cite{bethe79}. When the electron-to-baryon ratio has decreased sufficiently, and more neutron-rich and heavy nuclei are produced, $\beta^{-}$ decay also becomes important.

In order to understand the later stages of stellar evolution, the network calculations must be performed with accurate weak-interaction rates; pf-shell 
nuclei (A$\sim$40-65) are especially important. Fuller, Fowler and Newman first 
parameterized electron-capture rates \cite{fuller80,fuller82a,fuller82b,fuller85,ffn} by describing the Gamow-Teller strength in an independent-particle model. It is well-known, however, that the residual interactions between valence nucleons lead to a quenching and fragmentation of the strength \cite{BRO88}.  
For pf-shell nuclei, large-scale shell-model calculations \cite{caurier99,langanke00} were performed to estimate Gamow-Teller strengths, which were then used to predict weak-interaction rates in the stellar environment. Use of these new rates, instead of the ones calculated by Fuller, Fowler and Newman, have a strong effect on the predictions for the late evolution stages of stars \cite{heger01,heger01a,langanke03a}. It is, therefore, important that the theoretical strength distributions are reliable and in agreement with experimental data.
In this paper, we test theoretical predictions of the Gamow-Teller distribution in $^{58}$Co using the $^{58}$Ni(t,$^{3}$He) reaction at 115 MeV/nucleon. Here, the Gamow-Teller strength is defined so that the strength B(GT)=3 for the decay of a free neutron.

Charge-exchange reactions at intermediate energies (E$\gtrsim$100 MeV/nucleon) have been 
used widely to probe the spin-isospin response of nuclei (see e.g. \cite{osterfeld92,harakeh01}).  Both (p,n)-type 
($\Delta\text{T}_{z}=-1$) and (n,p)-type ($\Delta\text{T}_{z}=+1$) reactions 
have been employed. Gamow-Teller transition strengths ($\Delta\text{L}=0$, $\Delta\text{S}=1$) probed by nuclear reactions can be directly connected to weak-interaction strengths 
(for $\beta$ decay and electron capture) since the transitions, mediated through the $\sigma\tau_{\pm}$ operator, are between the same initial and final states in the two processes. 

Systematic studies of Gamow-Teller transition strengths were first 
performed at IUCF \cite{bainum80,goodman80,GAA81,RAP83,GAA85} using the 
(p,n) ($\Delta$T$_{z}=-1$) reaction.
It was shown \cite{taddeucci87} that there is a proportionality between the B(GT) values and the measured differential cross section at zero momentum transfer.  
This relationship between strength and cross section also provided the foundation for using other charge-exchange reactions, in particular the ($^{3}$He,t) reaction (see e.g. \cite{janecke93}), to measure Gamow-Teller strengths. Presently, the most extensive ($^{3}$He,t) program is 
performed at RCNP \cite{fujiwara96} (E($^{3}$He)=140-150 MeV/nucleon).  Using the 
dispersion-matching technique, an excitation-energy resolution of $\sim 30$ keV has 
been routinely achieved, see e.g. Refs. \cite{fujita02,FUJ03,FUJ04a,FUJ04b,FUJ05}.  

In the $\Delta$T$_{z}=+1$ direction, the (n,p) reaction has been used to measure Gamow-Teller strength distributions \cite{jackson88,ALF86,elkateb94}. Because resolutions are limited ($\sim$1 MeV), alternative probes have been developed. The (d,$^{2}$He) reaction was used at RIKEN \cite{okamura95}, Texas A\&M \cite{xu96}, and the KVI \cite{rakers02a}.
The most extensive program is performed at the latter institution \cite{GRE04,FRE04}, where resolutions in excitation energy of $\sim$130 keV at E(d)=85 MeV/nucleon are achieved.

In a recent paper \cite{zegers05}, we showed that the (t,$^{3}$He) reaction at 115 MeV/nucleon can also be used to extract Gamow-Teller strength distributions in the $\Delta$T$_{z}=+1$ direction. A study of the $^{26}$Mg(t,$^{3}$He) reaction was performed and combined with results from the $^{26}$Mg($^{3}$He,t) reaction. The details of the reaction mechanism were investigated. Gamow-Teller strengths extracted from (t,$^{3}$He) and ($^{3}$He,t) were compared with results from (d,$^{2}$He) and (p,n), respectively, displaying a good overall correspondence. Systematic errors in the extraction of Gamow-Teller strength using its proportionality to cross section at zero momentum transfer were studied and quantified. It was demonstrated that such errors (on the level of 10-20\% in the case of $^{26}$Mg) are largely due to interference between the $\Delta\text{L}=0$, $\Delta\text{S}=1$ Gamow-Teller amplitude and the $\Delta\text{L}=2$, $\Delta\text{S}=1$ amplitude mediated mainly by the tensor-$\tau$ component of the interaction. The interference can be constructive or destructive and it was found that the error in the strength averaged over many states became very small.

On the basis of these studies,
we now present the results from a measurement of the $^{58}$Ni(t,$^{3}$He) reaction and report the strength distribution in the pf-shell nucleus $^{58}$Co. 
In the past, the Gamow-Teller strength distribution has been obtained from $^{58}$Ni(n,p) \cite{elkateb94} and $^{58}$Ni(d,$^{2}$He) \cite{hagemann04,hagemann05} reactions. The reported Gamow-Teller strength distributions were inconsistent; the integrated strengths up to excitation energies of 10 MeV were similar, but the details of the strength distributions were different. 

In the stellar environment, electron-capture rates depend strongly on details of the Gamow-Teller distribution at low excitation energies due to phase-space effects. In Ref. \cite{hagemann04} it was shown that the calculated electron-capture rates based on the (d,$^{2}$He) and (n,p) results were significantly different ($\sim$20\%) for conditions in the core of a 25 solar-mass pre-supernova star following silicon depletion. A discrepancy in rates for one particular nucleus may not strongly affect the overall evolutionary track since typically electron-capture rates for about 5 nuclei are important at any stage \cite{heger01a}. It is important, however, to resolve the present ambiguity for the Gamow-Teller strength distribution in $^{58}$Co, since only a limited number of experiments are available to validate the shell-model calculations that are used to estimate the weak transition rates for a wide variety of nuclei.   

We note that $^{58}$Ni(t,$^{3}$He) experiments have been performed at beam energies of 8 MeV/nucleon \cite{ajzenberg-selove85} and 40 MeV/nucleon \cite{guillot04}. At these lower energies, systematic errors in the extraction of Gamow-Teller strengths can become large, due to the large magnitude of the tensor-$\tau$ interaction \cite{LOV81} and contributions from two-step processes \cite{fujiwara96}, and are thus not used to validate the Gamow-Teller strength distributions predicted by shell models. 

\section{Experiment}
\label{sec:experiment}

The production of a secondary triton beam at NSCL is described in detail in Refs. \cite{daito97,sherrill99}. The methods specific to the current data set include a primary $\alpha$ beam of 140 MeV/nucleon, accelerated in the K1200 cyclotron. The beam impinged on a thick Be production target (9.25 g/cm$^{2}$).  Tritons with a mean energy of 115 MeV/nucleon and an energy spread of 1\% were selected in the A1200 fragment 
separator and guided onto a target positioned at the object point of the S800 magnetic spectrometer \cite{bazin03}. The spectrometer is used to detect $^{3}$He particles from the (t,$^{3}$He) reaction and is operated in dispersion-matched mode to correct for the energy spread of the triton beam. It was positioned at 0$^{\circ}$ degrees and the solid angle covered in the measurements was about 20 msr.

The S800 focal-plane detector \cite{YUR99} was composed of two 2-dimensional cathode-readout 
drift detectors (CRDCs). They are used to measure the position and 
angles in the dispersive direction and non-dispersive direction at the focal plane.  The positions and angles were calibrated by inserting masks with holes and slits at known locations in front of the CRDCs and illuminating them with a $^3$He beam \cite{bazin03}.
For ray-tracing purposes, the ion-optical code COSY Infinity \cite{COSY} was used to calculate the ion-optical transfer matrix of the S800  spectrometer \cite{BER93} from the measured magnetic field maps. Matrix elements up to fifth order were used in the reconstruction of $\delta=(E-E_{0})/E_{0}$ ($E_{0}$ is the kinetic energy of the particle following the central-ray trajectory through the spectrometer and $E$ the energy of the measured particle). The track angles in the dispersive and non-dispersive directions were also obtained in the ray-tracing procedure and used to calculate the $^3$He scattering angle. The angular resolution was 10 mrad (FWHM). From these reconstructed parameters, the excitation energy was obtained from a missing-mass calculation with a resolution in the case of the $^{58}$Co spectrum of about 250 keV (FWHM). Angular distributions were extracted up to center-of-mass scattering angles of $5^{\circ}$.

Two thin plastic scintillation detectors (E1 and E2) were positioned behind the CRDCs.  The fast timing signal from the E1 scintillator served as the trigger for the data acquisition, as well as the start of the time-of-flight (TOF) measurement.  The stop signal for the TOF measurement was provided by the cyclotron radio frequency (RF). The light-output signal of E1 and E2 and the TOF signal were used to identify the $^3$He particles in the focal plane.  

The triton-beam intensity was monitored using an in-beam scintillator (IBS) 
placed 30 m upstream from the  target. The triton beam of 115 MeV/nucleon cannot be bent into the focal plane of the  
S800 spectrometer due to its high rigidity. The transmission from the 
IBS to the target was, therefore, measured by comparing the rates measured at 
the focal-plane scintillator and the IBS using a secondary $^3$He beam without a 
target and found to vary between 80 and 95\%. The secondary beam fills a large fraction of the available phase-space and, as a result, small changes in ion-optical settings or primary beam properties can lead to significant changes in transmission.  On average, the triton beam intensity at the target was about $1{\times}10^6$ pps.  

Besides the measurements using a 99.93\% isotopically enriched $^{58}$Ni target with a thickness of 7.61 mg/cm$^2$, measurements were also performed on a CH$_2$ target with a thickness of 6.72 mg/cm$^2$. The well-known Gamow-Teller transition to the $^{12}$B ground state provides a convenient way to check the procedures for extracting angular distributions. Because of the relatively strong kinematical correlation between momentum and angle for reactions on $^{12}$C, the excitation-energy resolutions ($\sim 300$ keV below $1^{\circ}$ to $\sim 450$ keV above $4^{\circ}$, respectively) were slightly worse than those obtained in the data taken on the $^{58}$Ni target.

In the experiment using the $^{58}$Ni target, the downstream CRDC detector was partially ($\sim$15\%) inefficient in the dispersive direction for $^{3}$He particles. Significantly reduced efficiency was confined to regions in the detector corresponding to excitation energies in $^{58}$Co higher than 10 MeV. Although momentum is mostly determined from the position measurement in the upstream CRDC (it is located near the true focal plane of the spectrometer), the scattering angle for events with missing downstream CRDC signal could not be well-determined. 
Since reconstruction of the angular distributions is important for extracting Gamow-Teller strengths from the spectra, the detailed analysis of the data taken on the $^{58}$Ni target was restricted to excitation energies below 10 MeV in $^{58}$Co.

\section{The $^{12}$C(\MakeTextLowercase{t},$^{3}$H\MakeTextLowercase{e})$^{12}$B reaction.}
\label{sec:analysisC}

\begin{figure}
\begin{center}
\includegraphics[scale=1.00]{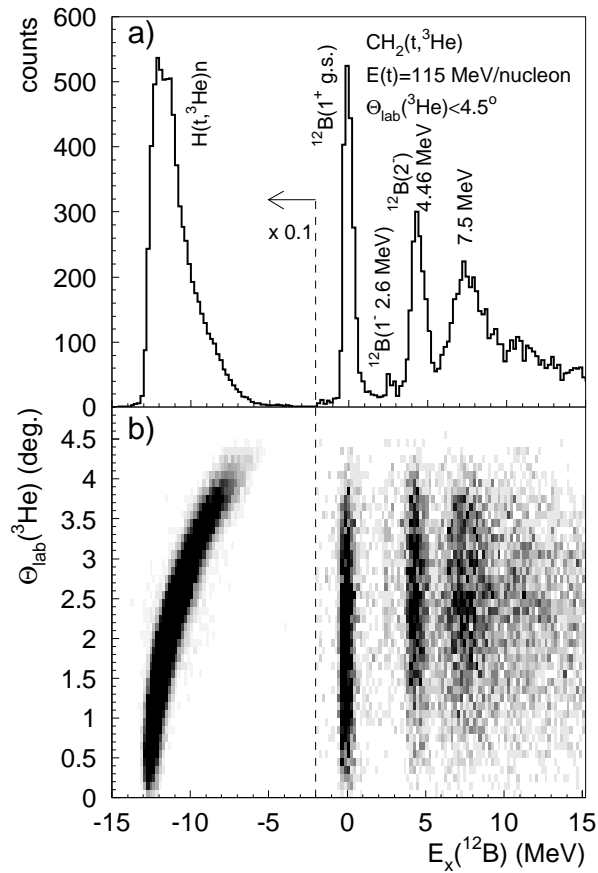}
\caption{a) Excitation-energy spectrum for the (t,$^{3}$He) reaction on a CH$_{2}$ target, for which kinematic corrections were made assuming a $^{12}$C target. The peak at `negative' excitation energies is due to reactions on hydrogen. The spectrum at $E_{x}<-2$ MeV has been down-scaled by a factor of 10. b) Scatter plot of the $^{3}$He laboratory scattering angle versus excitation energy in $^{12}$B.}
\label{12C}
\end{center}
\end{figure}

In Fig. \ref{12C}a, the excitation-energy spectrum from the CH$_{2}$(t,$^{3}$He) reaction is shown. In the missing-mass calculation, a $^{12}$C target is assumed and the events stemming from reactions on hydrogen in the target appear as an asymmetric peak at negative excitation energies (The Q-value for $^{12}$C(t,$^{3}$He)$^{12}$B(ground state) is -13.35 MeV and for H(t,$^{3}$He)n -0.764 MeV). Besides the $^{12}$B 1$^{+}$ ground state, other known states in the $^{12}$B spectrum are indicated in Fig. \ref{12C}a as well, along with their known $J^{\pi}$ assignments. The broad peak at around 7.5 MeV is due to several states, excited predominantly via dipole transitions ($J^{\pi}=1^{-},2^{-}$) but containing components with $J^{\pi}=3^{-}$ as well \cite{AJZ90}. Besides the ground state, two other $1^{+}$ states in $^{12}$B are known to exist (at 5.00 and 6.6 MeV \cite{AJZ90}) but were not identified in the current data set.    

In Fig. \ref{12C}b, a scatter plot of the laboratory scattering angle of the $^{3}$He versus the excitation energy is displayed. Because $^{12}$C is assumed to be the target in the missing mass calculation, the recoil energy for reactions on hydrogen in the target is underestimated, hence the remaining correlation between energy and angle. In the angular range under consideration ($\theta_{lab}<4.5^{\circ}$), the events from hydrogen in the target do not interfere with the analysis of the $^{12}$C data. 

\begin{figure}
\begin{center}
\includegraphics[scale=1.00]{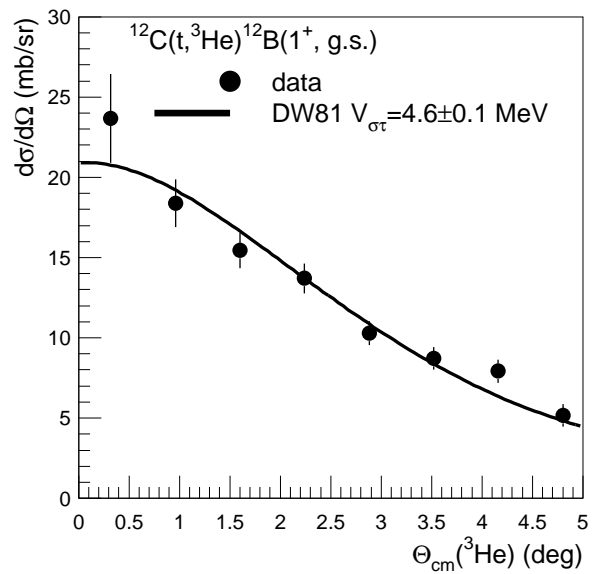}
\caption{The differential cross section of the $^{12}$C(t,$^{3}$He)$^{12}$B(g.s.) transition. The solid line is the result of the DWBA calculation (see text).}
\label{b12gs}
\end{center}
\end{figure}

The code DW81 \cite{schaeffer70} was used to calculate differential cross sections for the transition to the $^{12}$B $J^{\pi}=1^{+}$ ground state. The B(GT) for this transition equals 0.990$\pm$0.002 \cite{CHO93}, which is calculated from measured $\beta$ decay log$ft$ values \cite{AJZ90}.
 The optical-potential parameters for the outgoing $^{3}$He particle were obtained from elastic scattering of $^3$He on $^{12}$C at 150 MeV/nucleon 
\cite{kamiya03}. Following Refs. \cite{grasdijk,vanderwerf89}, the optical potential parameters for the triton were 
obtained by scaling the $^3$He potential depths by 0.85 \cite{vanderwerf89}, while keeping the radii and diffusenesses constant. The one-body-transition-densities (OBTDs)  were 
calculated using the code OXBASH \cite{OXBA} with the CKII interaction \cite{cohen67} 
in the p model space. The Gamow-Teller strength obtained in this calculation for the transition to the $^{12}$B ground state (B(GT)=0.98) is close to the experimental value from $\beta$ decay.
Wave functions of the target and residual nucleus were calculated in a Woods-Saxon potential. Single-particle binding energies were determined in the code OXBASH \cite{OXBA} using the Skyrme SK20 interaction \cite{BRO98}. The projectile-target interaction was described by an 
effective $^3$He-nucleon interaction with spin-isospin (V$_{\sigma\tau}$), isospin (V$_{\tau}$) and
tensor-isospin (V$_{T\tau}$) components 
\cite{vanderwerf89}. The former two are described by single Yukawa potentials, for which the ranges are fixed to that of the one-pion exchange potential, $R=1.414$ fm. The tensor-isospin term is represented by a potential of the form $r^{2}\times\text{Yukawa}$ (with range $R=0.878$ fm), multiplied by the tensor operator $S_{12}=\frac{(\mathbf{\sigma_{1}\cdot \vec{r}})(\mathbf{\vec{\sigma_{2}}\cdot \vec{r}})}{r^{2}}-\mathbf{\vec{\sigma_{1}}\cdot\vec{\sigma_{2}}}$. In principle, a spin-orbit term (V$_{LS\tau}$) should also be included, but it is usually set to zero since it was predicted \cite{LOV81} and confirmed by experiment \cite{grasdijk,schaeffer71} that it hardly contributes at low momentum transfers. Since spin-transfer is required, only V$_{\sigma\tau}$ and V$_{T\tau}$ are important for Gamow-Teller transitions.  The ratio V$_{\sigma\tau}$/V$_{T\tau}$ was fixed to $1.0$ following Ref.
\cite{zegers03}.  

In Fig. \ref{b12gs}, the differential cross section for the transition to the $^{12}$B ground state is compared with the DWBA calculations. The error bars reflect statistical uncertainty only. In the comparison between data and theory, the magnitude of V$_{\sigma\tau}$ was determined by a fit to the data. Since the ratio V$_{\sigma\tau}$/V$_{T\tau}$ was fixed, V$_{T\tau}$ was adjusted accordingly. The fitted value of V$_{\sigma\tau}$ was $4.6\pm 0.1$ MeV ($\chi^{2}/N_{d.o.f.}=1.3$). The good correspondence between experiment and theory confirms the findings from the experiment on $^{26}$Mg \cite{zegers05}, namely that the DWBA calculations can be used to predict reliably angular distributions. For the analysis of the $^{58}$Ni data, the value of V$_{\tau}$ was set to 1.6 MeV using the known ratio of V$_{\sigma\tau}$/V$_{\tau}=2.9$ \cite{AKI95}.

\section{The $^{58}$N\MakeTextLowercase{i}(\MakeTextLowercase{t},$^{3}$H\MakeTextLowercase{e})$^{58}$C\MakeTextLowercase{o} reaction.}
\label{sec:analysisN}

\begin{figure}
\begin{center}
\includegraphics[scale=1.00]{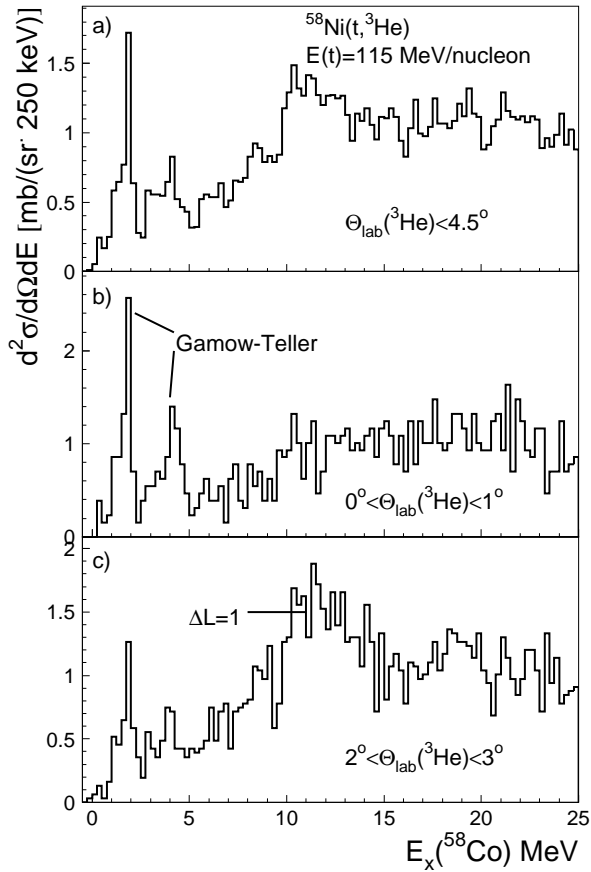}
\caption{a) The $^{58}$Co excitation-energy measured via the $^{58}$Ni(t,$^3$He) reaction at 115 MeV/nucleon integrated over the full measured solid angle. b) The same reaction, but gated on events with ${\theta}_{lab}=$0$^{\circ}$-1$^{\circ}$. The two indicated peaks can be identified as Gamow-Teller transitions because of their maxima at forward scattering angles. c) The same reaction, but gated on events with ${\theta}_{lab}=$2$^{\circ}$-3$^{\circ}$. The indicated broad structure broad is likely due to dipole transitions, identified by its maximum in this angular range. }
\label{co58spectrum}
\end{center}
\end{figure}

In Fig. \ref{co58spectrum}a, the $^{58}$Ni(t,$^{3}$He) spectrum is shown integrated over the full angular range covered in the present experiment. A binning of 250 keV is applied, corresponding to the energy resolution. The Q-value for the transition to the $J^{\pi}=2^{+}$ ground state of $^{58}$Co is -0.363 MeV, less negative than that of the H(t,$^{3}$He)n reaction. By examining the correlation between scattering angle and $^{3}$He energy, no significant contributions from events due to the H(t,$^{3}$He)n reaction to the $^{58}$Co spectrum were found.
   
\begin{figure}
\begin{center}
\includegraphics[scale=0.90]{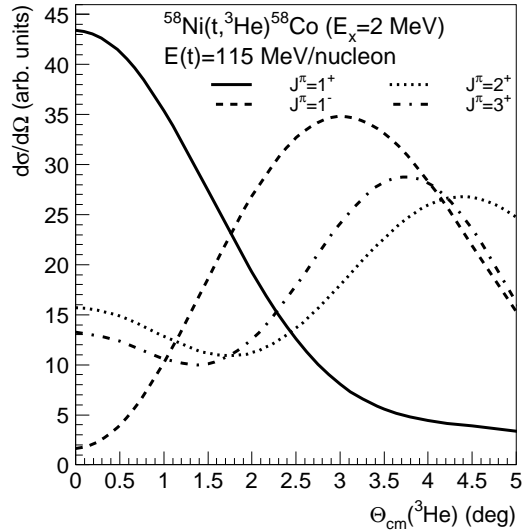}
\caption{Differential cross sections calculated in DWBA for $^{58}$Ni(t,$^{3}$He) transitions to states in $^{58}$Co with different $J^{\pi}$. In all cases, an excitation energy of 2 MeV in $^{58}$Co is assumed and the normalizations are adjusted so that the cross sections are of the same order of magnitude for easy comparison.}
\label{ang}
\end{center}
\end{figure}

Transitions of various multipolarities contribute to the spectrum shown in Fig. \ref{co58spectrum}a. Whereas transitions with $\Delta L=0$ are associated with differential cross sections that are strongly forward-peaked, differential cross sections of dipole transitions ($\Delta L=1$) peak at about $3^{\circ}$ and transitions due to higher angular-momentum transfer are relatively flat at very forward angles and have a maximum at $\sim 4^{\circ}$. In Fig. \ref{ang} angular distributions calculated in DWBA for several relevant transitions to states of different multipolarity are compared. The details of these calculations are discussed below.

To illustrate the method of using the angular distributions to identify different types of transitions in the data, the excitation energy spectra for $0^{\circ}<\theta_{\text{lab}}(^{3}\text{He})<1^{\circ}$ and $2^{\circ}<\theta_{\text{lab}}(^{3}\text{He})<3^{\circ}$ are displayed in Figs. \ref{co58spectrum}b and \ref{co58spectrum}c, respectively. Just below $E_{x}=2$ MeV and at about $E_{x}=4$ MeV two strongly forward-peaked transitions can be identified (as indicated in Fig. \ref{co58spectrum}b), revealing the presence of transitions with $\Delta L=0$. For $7<E_{x}<15$ MeV, the differential cross section rises at backward angles, revealing the presence of dipole contributions presumably due to the various components of the isovector spin giant dipole resonance ($\Delta L=1$, $\Delta S=1$, $J^{\pi}=0^{-},1^{-},2^{-}$) and its non-spin-flip partner, the isovector giant dipole resonance ($\Delta L=1$, $\Delta S=0$, $J^{\pi}=1^{-}$). The main bump is indicated in Fig. \ref{co58spectrum}c. Note that even though for about 15\% of the data above $E_{x}>10$ MeV the angle was not well determined in the analysis (see Section \ref{sec:experiment}) such gross features remain visible. 

Since transitions with different angular momentum transfer are not necessarily separated in energy, a multipole decomposition analysis (MDA) must be performed to extract the strength distributions as discussed below.  
In the $T_{z}=+1$ direction, the possible $\Delta L=0$ transitions are to Gamow-Teller states and the $2\hbar\omega$ isovector (spin-flip) giant monopole resonances \footnote{In the $\Delta T_{z}=-1$ direction, the same is nearly true, except for the presence of the isobaric analog state ($\Delta L=0$, $\Delta S=0$).}. The latter resonances only contribute to the $^{58}$Co spectrum at $E_{x}\gtrsim 10$ MeV \cite{AUE83,AUE84,guillot04} and peak at about $E_{x}=25$ MeV  \footnote{In Refs. \cite{AUE83,AUE84} calculations are performed for $^{60}$Ni and not $^{58}$Ni. The general properties of the isovector giant monopole resonances are not expected to chance rapidly as a function of mass number, however.}. Therefore, extraction of the $\Delta L=0$ component in the region $E_{x}<10$ MeV based on angular distributions is a very selective tool for identifying the contributions due to Gamow-Teller transitions.       

A MDA was performed for each 250-keV wide bin, by fitting the experimental differential cross section to a linear superposition of theoretically predicted angular distributions with the code DW81 \cite{schaeffer70} for various types of transitions. The optical potential parameters for the $^{58}$Ni target were taken from fitting $^3$He elastic scattering data at 
150 MeV/nucleon \cite{kamiya03}.  Wave functions projected on a complete $1p$-$1h$ basis were calculated in a normal-modes procedure \cite{HOF95,zegers01} using the code NORMOD \cite{WER91}. In this formalism, the full strength (i.e. 100\% of the non-energy-weighted sum rule) associated with the single-particle multipole operators is exhausted. The calculated cross sections, therefore, do not relate to any particular state in the spectrum and the main purpose of the calculations was only to determine the shape of the angular distribution used in the MDA. The DWBA calculations were performed for each excitation-energy bin separately to account for the increase in momentum transfer with increasing excitation energy at a fixed scattering angle.

\begin{figure}
\begin{center}
\includegraphics[scale=0.90]{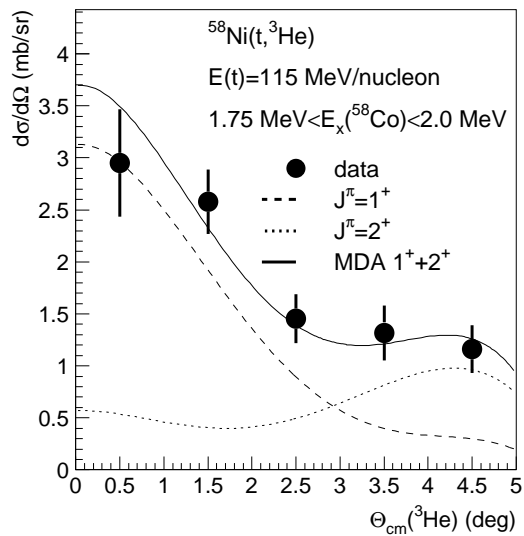}
\caption{The differential cross-sections for the energy bin $1.75<E_{x}(^{58}\text{Co})<2.0$ MeV and the result of the MDA (solid line) using a linear combination of $1^{+}$ (dashed line) and $2^{+}$ (dotted line) components. The error bars in the data are of statistical nature only.}
\label{mda}
\end{center}
\end{figure}

The statistical accuracy of the data was limited, and only five 1$^{\circ}$-wide angular bins between 0$^{\circ}$ and 
5$^{\circ}$ in the center-of-mass were used. With a limited number of data points, a fit with multiple components is difficult. Therefore, we used only two multipole components per excitation-energy bin: a Gamow-Teller component and one component with larger angular-momentum transfer. The choice for the second component was made based on the excitation-energy of the bin. In the region below  $E_{x}=5$ MeV, states with $J^{\pi}=1^+,2^+,3^+,4^+,5^+$ are known to reside and contributions from dipole transitions are small \cite{BHA97}. At beam energies above 100 MeV/nucleon and at forward angles, transitions with large orbital angular momentum transfers are strongly suppressed and states with $J^{\pi}=4^+,5^+$ are hardly excited and can safely be ignored \cite{zegers05}. The transitions to the $J^{\pi}=2^+,3^+$ states have similar distributions (see Fig. \ref{ang}). As a result, differences between fits with a Gamow-Teller component and either a $J^{\pi}=2^+$ or a $J^{\pi}=3^+$ component resulted in differences for the extracted Gamow-Teller cross sections that were smaller than the statistical error. Therefore, for extracting Gamow-Teller strength in the excitation-energy region below 5 MeV, the MDA was performed using a Gamow-Teller and $J^{\pi}=2^+$ component only. As an example of the procedure, the result of the MDA in the energy bin from 1.75 to 2 MeV is shown in Fig. \ref{mda}. A systematic error of 5\% was assigned to the extracted Gamow-Teller strength in each bin based on the differences between fits with a $J^{\pi}=2^+$ or a $J^{\pi}=3^+$ component.

For the region $5<E_{x}<10$ MeV, the situation is more complicated, since dipole transitions will play an increasingly larger role at higher excitation energies. It was found that the fits to the angular distributions had consistently lower $\chi^{2}$ values per degree of freedom when using a dipole component \footnote{Relative to spin-transfer transitions, the non-spin-transfer component is expected to be weak \cite{LOV81} at this beam energy. In addition, owing to the small repulsive spin-orbit interaction for configurations leading to excitation of the $J^{\pi}=2^{-}$ component, relative to excitation of the $1^{-}$ and $0^{-}$ components of the spin-flip dipole resonance \cite{BER81}, it is most likely that low-lying dipole strength is associated with $2^{-}$ states. Therefore, dipole transitions were assumed to populate $2^{-}$ states only, although calculations assuming population of $J^{\pi}=0^{-}$ or $1^{-}$ gave nearly identical results.} in the MDA in addition to a Gamow-Teller component, compared to using transitions leading to states with $J^{\pi}=2^+$ or $J^{\pi}=3^+$. Irrespective of the choice for the component used in addition to the contribution from the Gamow-Teller component, the extracted Gamow-Teller cross section did not change significantly (i.e. by more than statistical error bars). A systematic error of 10\% of the extracted Gamow-Teller in each bin was assigned based on the differences between fits with a $J^{\pi}=2^+,3^{+}$ or a dipole component.

\section{Extraction of Gamow-Teller strengths.}
\label{sec:extract}
The next step in the analysis is to convert the extracted Gamow-Teller contributions in the excitation-energy spectrum to strength. This is done by using the proportionality between the cross section at zero momentum transfer ($\frac{d\sigma}{d\Omega}(q=0)$) and Gamow-Teller strength (B(GT)) as derived in Eikonal approximation \cite{goodman80,taddeucci87}: 
\begin{equation}
\frac{d{\sigma}}{d{\Omega}}(q=0) = KN^D|J_{\sigma\tau}|^2B(GT).
\label{eq:general}
\end{equation}
Here, $K=\frac{E_{i}E_{f}}{(\hbar^{2}c^{2}\pi)^{2}}$, where $E_{i(f)}$ is the reduced energy in the incoming (outgoing) channel. $|J_{\sigma\tau}|^2$ is the volume integral of the $\sigma\tau$ component of the projectile-target interaction.  The distortion factor 
$N^D$ is defined by \cite{taddeucci87}
\begin{equation}
N^D = \frac{\frac{d\sigma}{d\Omega}_{DWBA}(q=0)}{\frac{d\sigma}{d\Omega}_{PWBA}(q=0)},
\label{eq:distortionf}
\end{equation}
and is determined using the DWBA code. For the plane-wave (PWBA) calculation, the depth of the optical potentials and the charges of the target and residual nuclei are set to zero. The factor $KN^{D}|J_{\sigma\tau}|^{2}$ is referred to as the unit cross section $\hat{\sigma}$.

Application of Eq. \ref{eq:general} to calculate B(GT), requires the knowledge of the experimental cross section at zero momentum transfer (i.e. requiring that both the Q-value ($Q$) of the transition and the scattering angle are zero), which is obtained from:  
\begin{equation}
\label{eq:extra}
\frac{d\sigma}{d\Omega}(q=0)=\left[\frac{\frac{d\sigma}{d\Omega}(q=0)}{\frac{d\sigma}{d\Omega}(Q,0^{\circ})}\right]_{\text{DWBA}}\times\left[\frac{d\sigma}{d\Omega}(Q,0^{\circ})\right]_{\text{exp}}.
\end{equation}
In this equation, `$\text{DWBA}$' refers to values calculated in the DWBA code. The experimental cross section at $\theta=0^{\circ}$ and its error are taken from the fitted Gamow-Teller angular distribution in the MDA.   

In practice, unit cross sections are determined using a transition for which B(GT) is known from $\beta$ decay; we apply the same method here, albeit indirectly. As with the procedure used in the analysis of the $^{58}$Ni(d,$^{2}$He)$^{58}$Co experiment \cite{hagemann04,hagemann05} the unit cross section was determined from the transition to the strongest $1^{+}$ state at $E_{x}=1.87$ MeV present in the spectrum. The analog of this transition has been studied with the $^{58}$Ni($^{3}$He,t)$^{58}$Cu reaction at 140 MeV/nucleon \cite{Fuj02,HFuj02}. Under the assumption of isospin symmetry, the difference in B(GT) between the analogs is only the square of an isospin Clebsch-Gordan coefficient. In the case of the $^{58}$Ni target, the strength is a factor of 6 larger for the $\Delta T_{z}=+1$ transition than for the $\Delta T_{z}=-1$ transition. The Gamow-Teller strength for the ground state transition from  $^{58}$Ni to $^{58}$Cu was determined from the measured log$ft$ value for $\beta$ decay \cite{BHA97}. 
This yields a Gamow-Teller strength of the $1^{+}$ state at 1.87 MeV in $^{58}$Co of $0.72\pm0.05$. Another $1^{+}$ state is known to be located at $E_{x}=1.73$ MeV in $^{58}$Co \cite{BHA97}. Its B(GT) was determined to be $0.17\pm0.04$ \cite{HFuj02}.  Since in the present (t,$^{3}$He) experiment this state was not separated from the one at $E_{x}=1.87$ MeV, the unit cross section was determined using the sum of these states. This calibration procedure has a systematic error that will be discussed in more detail in Section \ref{sec:sys}. Once the unit cross section was determined, it was used to convert the cross sections at $0^{\circ}$ obtained from the MDA analysis in each energy bin to strengths, using Eqs. \ref{eq:general}-\ref{eq:extra}.

\section{Results and comparison with theory and previous data.}
\label{sec:results}

\begin{figure}
\begin{center}
\includegraphics[scale=1.0]{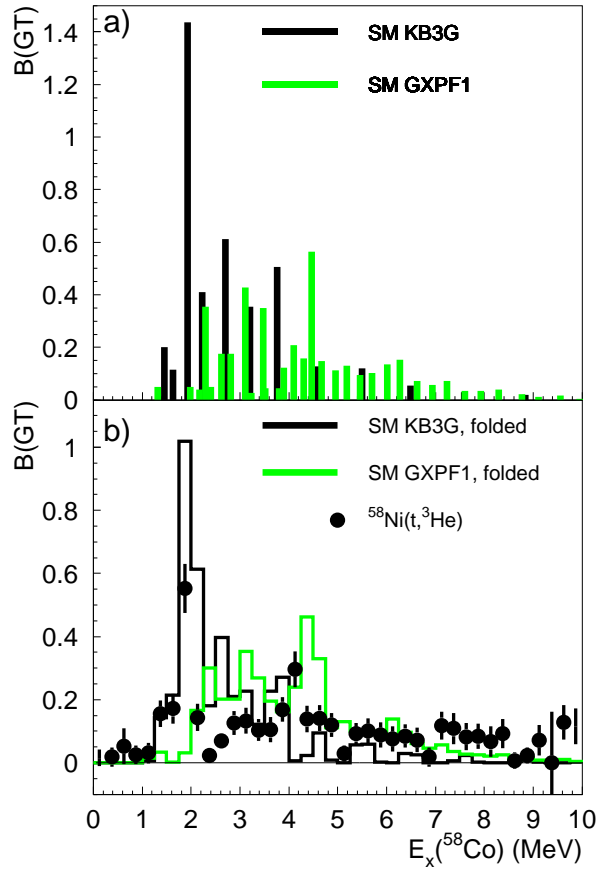}
\caption{a) Large-scale shell-model calculations for the Gamow-Teller strength distribution in $^{58}$Co using the KB3G \cite{POV01} interaction \cite{hagemann04,hagemann05,MP} (black) and GXPF1 \cite{HON02,HON04} interaction \cite{hagemann05} (green). b) Gamow-Teller strength distribution extracted from the $^{58}$Ni(t,$^{3}$He) data compared with the theoretical distributions that were folded with the experimental resolution (250 keV) and binned in the same manner as the data.}
\label{bgt}
\end{center}
\end{figure}

In Fig. \ref{bgt} the extracted Gamow-Teller distribution for excitation energies between 0 MeV and 10 MeV in $^{58}$Co from the $^{58}$Ni(t,$^{3}$He) experiment is shown and compared with theory. In both theoretical calculations, a quenching factor of $(0.74)^{2}$ \cite{MAR96} is applied. The different interactions result in significantly different strength distributions. 

To facilitate the comparison between the data and theory in Fig. \ref{bgt}b, the calculations shown in Fig. \ref{bgt}a were folded with the experimental resolution and binned in the same manner as the data. The calculation with the KB3G interaction does relatively well in describing qualitatively the experimental strength distribution up to about $E_{x}=4$ MeV, but is too large in magnitude. Above $E_{x}=4$ MeV, too little strength is predicted. The calculation with the GXPF1 interaction does not reproduce the strongest state at 1.87 MeV, but predicts more strength at higher excitation energies and is in that respect more consistent with the data.

Next, we compare the results from the $^{58}$Ni(t,$^{3}$He) experiment with previous experimental results from $^{58}$Ni(n,p) \cite{elkateb94} and $^{58}$Ni(d,$^2$He) \cite{hagemann04,hagemann05}.
Better energy resolution is achieved in the latter experiment (130 keV). The energy resolution in the (n,p) experiment was 1.3 MeV. Therefore, when comparing the various results, the resolution must be taken into account, as in the comparison between the $^{58}$Ni(n,p) and $^{58}$Ni(d,$^2$He) results made in Refs. \cite{hagemann04,hagemann05}. However, in those Refs. the bins for the two data sets were shifted by 0.5 MeV relative to each other, which made a direct comparison somewhat difficult. Therefore, we performed the comparison between the (d,$^{2}$He) and (t,$^{3}$He) results and the comparison between the (n,p) and (t,$^{3}$He) results separately. 

\begin{figure}
\begin{center}
\includegraphics[scale=1.00]{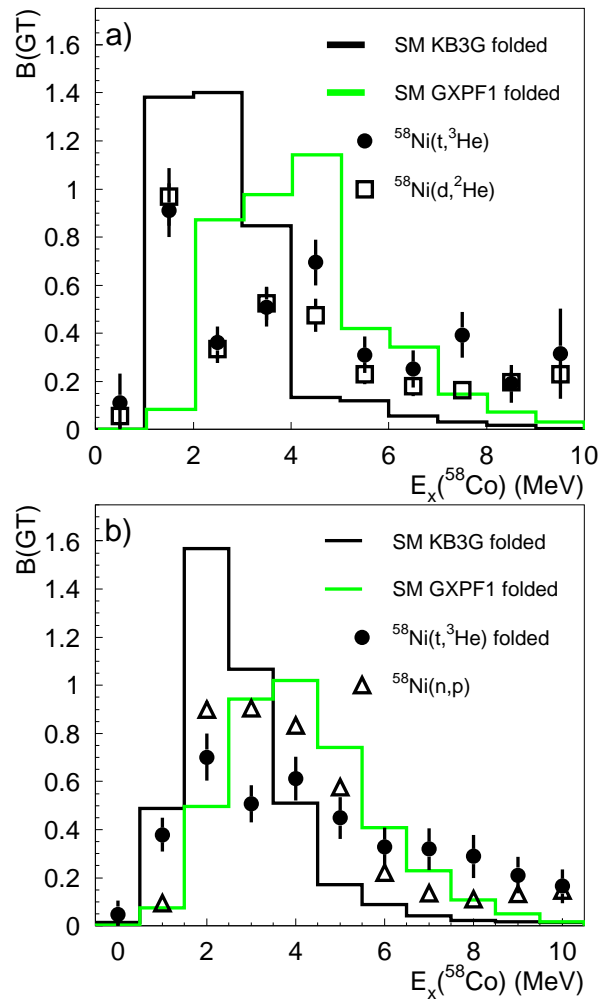}
\caption{a) Comparison of the results of the $^{58}$Ni(d,$^{2}$He) and $^{58}$Ni(t,$^{3}$He) experiments and the theoretical predictions. A binning of 1 MeV was applied and the theory was folded with the experimental resolution of the (t,$^{3}$He) experiment (250 keV) before binning. b) Comparison of the results of the $^{58}$Ni(n,p) and $^{58}$Ni(t,$^{3}$He) experiments and the theoretical predictions. A binning of 1 MeV was applied. Note the 0.5 MeV shift relative to a). The (t,$^{3}$He) data set and theory were folded with the experimental resolution of the (n,p) experiment (1.3 MeV) before binning.}
\label{fig8}
\end{center}
\end{figure}

In Fig. \ref{fig8}a, the (d,$^{2}$He) and (t,$^{3}$He) results are compared. Results from $^{58}$Ni(d,$^2$He) were directly taken from Table 2 in Ref. \cite{hagemann04}. For the $^{58}$Ni(t,$^3$He) data, two alternative methods to extract the Gamow-Teller strength distribution in 1-MeV bins were used: i) The MDA analysis was performed in each 1-MeV bin and ii) the extracted B(GT)s from four consecutive 250-keV bins were added. The differences between methods i) and ii) were smaller than the statistical errors, and in Fig. \ref{fig8}a the results from method ii) are shown. Since the energy resolution in both data sets is much smaller than the bin size, the difference in energy resolution hardly affects the representation and was not corrected for. The (d,$^{2}$He) and (t,$^{3}$He) data sets are consistent. The $\chi^{2}/N_{\text{d.o.f.}}=1.21$ with $N_{\text{d.o.f}}=9$, i.e. well within a 95\% confidence interval (since the absolute scale of the distribution was determined in the same manner for the two data sets, $N_{\text{d.o.f}}=9$ instead of 10). Also included in the figure are the theoretical distributions, folded with the experimental resolution achieved in the (t,$^{3}$He) experiment and binned in the same manner as the data. 

In Fig. \ref{fig8}b, the (n,p) and (t,$^{3}$He) results are compared. Again, a bin size of 1 MeV is used, but shifted by 0.5 MeV relative to Fig. \ref{fig8}a. Since the (n,p) data had a much lower energy resolution, the resolution of (t,$^{3}$He) data was artificially reduced to that of the (n,p) data. The two different theoretical calculations were also folded with the resolution from the (n,p) experiment and binned according to the data. As can be seen from the (t,$^{3}$He) strength distribution in Fig. \ref{fig8}b, the folding and choice of binning almost hides the presence of a strong peak at $E_{x}=1.87$ MeV, and it is, therefore, understandable that it was not seen in the (n,p) data. 
However, the (t,$^{3}$He) and (n,p) data sets are not consistent ($\chi^{2}/N_{\text{d.o.f.}}=6.0$ with $N_{\text{d.o.f}}=10$, i.e. well outside the 99.5\% confidence interval). The unit cross section in the (n,p) experiment \cite{elkateb94} was obtained by studying its systematic behavior in nuclei of similar mass. Although this could lead to a systematic discrepancy with the (t,$^{3}$He) results, we note that rescaling either of the data sets does not improve the consistency, since for $E_{x}<5$ MeV ($E_{x}>5$ MeV) the (n,p) strengths are larger (smaller) than the (t,$^{3}$He) strengths. It is important to note that, if the data sets with relatively good resolutions (from (d,$^{2}$He) and (t,$^{3}$He)) were not available, one might have concluded that the theoretical calculation with the GXPF1 interaction is favored for predicting the Gamow-Teller distribution since, except for a minor shift in excitation energy, the match with the (n,p) results is good. But after comparison with the (d,$^{2}$He) and (t,$^{3}$He) results it becomes clear that the strong state at $E_{x}=1.87$ MeV is missed when using the GXPF1 interaction. Reiterating a conclusion from Ref. \cite{hagemann04}, it is necessary to obtain data with good resolution to test shell-model calculations in detail.

Although the two theoretical calculations and the three data sets have different strength distributions, the summed strengths up to $E_{x}=10$ MeV are very similar: $\sum_{\text{KB3G}}\text{[B(GT)]}=4.0$, $\sum_{\text{GXPF1}}\text{[B(GT)]}=4.1$ and $\sum_{(\text{t},^{3}\text{He})}\text{[B(GT)]}=4.1\pm0.3$, $\sum_{(\text{d},^{2}\text{He})}\text{[B(GT)]}=3.4\pm0.2$ and 
$\sum_{(\text{n},\text{p})}\text{[B(GT)]}=4.1\pm0.1$ (the errors in the summed strengths for the three data sets are statistical only). 

\section{The normalization of B(GT)}
\label{sec:sys}
As discussed in Section \ref{sec:extract}, the Gamow-Teller strength distribution in $^{58}$Co was normalized using known strengths for the analog of the $1^{+}$ states at 1.73 MeV and 1.87 MeV studied with the $^{58}$Ni($^{3}$He,t) reaction at 140 MeV/nucleon \cite{Fuj02,HFuj02}. In turn, the B(GT) of these analog states, located at 10.60 MeV and 10.82 MeV in $^{58}$Cu, were calibrated with the B(GT) known from $\beta$ decay of the ground state transition $^{58}$Ni(g.s.)$\rightarrow^{58}$Cu(g.s.). Transitions from $^{58}$Ni to $^{58}$Cu have also been studied with the (p,n) reaction \cite{RAP83} at 120 MeV and 160 MeV. After determining the unit cross section for the ground-state transition, in both the ($^{3}$He,t) and (p,n) data sets the B(GT) for the second $1^{+}$ state at 1.05 MeV was extracted as well. However, the ratios of B(GT)s for the transition to the ground state and to the second $1^{+}$ state in $^{58}$Cu are very different for the three data sets. This is shown in Table \ref{tab:cu}. In the (p,n) experiments, the $^{58}$Cu ground state was not separated from the isobaric analog state ($J^{\pi}=0^{+}$) at $E_{x}$($^{58}$Cu)=0.20 MeV, which could have led to a systematic error. Perhaps, the difference between the ratios between the ratios in Table \ref{tab:cu} for the two (p,n) experiments is an indication of that. However, this cannot fully explain the large difference between the two (p,n) and the ($^{3}$He,t) result.

\begin{table*}
\caption{\label{tab:cu} Comparison between Gamow-Teller strengths extracted for the ground state and second $1^{+}$ state in $^{58}$Cu via $^{58}$Ni(p,n) at 120 MeV and 160 MeV \cite{RAP83} and $^{58}$Ni($^{3}$He,t) \cite{Fuj02,HFuj02}.}
\begin{ruledtabular}
\begin{tabular}{lccc}
& $^{58}$Ni(p,n) 120 MeV \footnotemark[1] & $^{58}$Ni(p,n) 160 MeV\footnotemark[1] & $^{58}$Ni($^{3}$He,t) 140 MeV/nucleon \\ \hline
B(GT) $^{58}$Cu (1$^{+}$ g.s.) & 0.165\footnotemark[2] & 0.165\footnotemark[2] & 0.155$\pm$0.001\footnotemark[2] \\
B(GT) $^{58}$Cu (1$^{+}$ 1.05 MeV) & 0.5 & 0.4 & 0.265$\pm$0.013 \\ \hline
$\frac{\text{B(GT)} ^{58}\text{Cu} (1^{+} \text{g.s.})}{\text{B(GT)} ^{58}\text{Cu} (1^{+} \text{1.05 MeV})}$ & 0.33 & 0.41 & 0.58 \\
\end{tabular}
\end{ruledtabular}
\footnotetext[1]{No explicit error bars for B(GT) values are given in Ref. \cite{RAP83}. The authors note that the error in the absolute cross sections for each state are $\sim15$\%}
\footnotetext[2]{Fixed to the known B(GT) from $\beta$ decay \cite{BHA97}. The small difference between the value for the (p,n) reactions and the ($^{3}$He,t) reaction stems from the constants used when converting the log$ft$ value to B(GT).} 
\end{table*}

In Ref. \cite{HON05}, the Gamow-Teller strength distributions in $^{58}$Cu calculated using the KB3G and GXPF1 interactions were discussed. When using the GXPF1 interaction, the overall strength distribution was well reproduced, but the B(GT) for the transition to the ground $1^{+}$ state at $E_{x}$=1.05 MeV was too low compared to experiment. With the KB3G interaction, the B(GT) for that state was better reproduced, but a group of $1^{+}$ states experimentally found around $E_{x}$=3.5 MeV was absent.  

In a simple independent-particle shell model, the neutrons and protons in $^{58}$Ni fill all orbitals up to $f_{7/2}$. The two remaining neutrons populate the $p_{3/2}$ and $f_{5/2}$ orbitals. Therefore, relatively strong contributions from $\nu p_{3/2}$-$\pi f_{5/2}$ and $\nu f_{5/2}$-$\pi p_{3/2}$ particle-hole components are to be expected for excitations of the lowest lying $1^{+}$ states in $^{58}$Cu, even though these components are purely $\Delta L=2$, $\Delta S=1$ nature (i.e. not Gamow-Teller). As discussed in Ref. \cite{HON05}, in more realistic models using the GXPF1 and KB3G interactions, this is still largely true: transitions to the $1^{+}$ ground state and excited state at $E_{x}$=1.05 MeV have strong contributions from the $p$-orbit. 
The components with $\Delta L=2$, $\Delta S=1$ tend to break the proportionality between cross section at zero momentum transfer and B(GT) in Eq. \ref{eq:general} \cite{taddeucci87,zegers05}. Since the effect of the interference between $\Delta L=2$, $\Delta S=1$ and $\Delta L=0$, $\Delta S=1$ amplitudes on the cross section depends on the reaction and beam energy, it could possibly explain the discrepancies between the ($^{3}$He,t) and (p,n) results shown in Table \ref{tab:cu}. 

In the $\beta^{+}$, i.e. (n,p) direction, contributions from $\Delta L=2$, $\Delta S=1$ components to the Gamow-Teller transitions are relatively small in the case of the $^{58}$Ni target, since (in an independent-particle picture) the protons do not fill the $p_{3/2}$ and $f_{5/2}$ orbitals. 

To estimate the proportionality-breaking effects we performed cross-section calculations for the $^{58}$Ni(p,n) and $^{58}$Ni($^{3}$He,t) reactions for the transitions to the $1^{+}$ ground state and excited state at $E_{x}$=1.05 MeV in $^{58}$Cu. The beam energy was set to 140 MeV/nucleon, which is between the beam energies of the two available (p,n) measurements and equal to the energy used in the ($^{3}$He,t) experiment . 

For the $^{58}$Ni(p,n) reaction, the code DW81 \cite{schaeffer70} was used. The Love-Franey effective interaction at 140 MeV \cite{LOV81,LOV85} was employed and exchange effects were treated exactly. OBTDs for the transitions to the relevant $1^{+}$ states in $^{58}$Cu were taken from shell-model calculations in which at most 5 nucleons are allowed to be excited from the $f_{5/2}$ orbit to higher-lying orbits \cite{HON05,HON06}. To test the dependence of the results on the interaction, two sets of OBTDs were used, calculated with the KB3G and the GXPF1 interactions. Radial wave functions of the target and residual nuclei were calculated using a Woods-Saxon potential. Single-particle binding energies of the particles were determined in OXBASH \cite{OXBA} using the Skyrme SK20 interaction \cite{BRO98}. Optical-model parameters were calculated following the procedure in Ref. \cite{NAD81} which includes Coulomb corrected isovector terms to account for the differences between the incoming (proton plus $^{58}$Ni) and outgoing (neutron plus $^{58}$Cu) channel.  

In order to perform a precise comparison with the (p,n) results, the calculations for the $^{58}$Ni($^{3}$He,t) reaction need to be performed with the same effective interaction (i.e. the Love-Franey interaction \cite{LOV81,LOV85} instead of the effective $^{3}$He-nucleon interaction used above). Therefore, the code FOLD \cite{FOLD} was employed (for more details see Ref. \cite{zegers05}), in which the Love-Franey interaction is double-folded over the transition densities. Exchange is treated in the short-range approximation described in Ref. \cite{LOV81}. OBTDs for the target-residual nucleus system are the same as those used in the (p,n) calculation \cite{HON06}. For $^{3}$He and $^{3}$H, densities were obtained from Variational Monte-Carlo results \cite{WIR05}. Optical-model parameters for the incoming $^{3}$He channel were taken from Ref. \cite{kamiya03}. Following the analysis in Ref. \cite{vanderwerf89}, the potential-well depths were scaled with a factor 0.85 for the outgoing triton channel.  

All cross-section calculations were performed at zero-momentum transfer by setting the Q-value of the reaction to zero and using the cross section at $0^{\circ}$ scattering angle. For each transition and for each set of OBTDs, three calculations were performed: i) a plane-wave (PW) calculation, ii) a distorted-wave (DW) calculation and iii) a distorted-wave calculation in which the tensor-$\tau$ amplitudes of the Love-Franey interaction were set to zero (DW$_{T\tau=0}$). For each calculation, a unit cross section was calculated: $\hat{\sigma}=\frac{d\sigma}{d\Omega}(q=0)/\text{B(GT)}$ and subsequently the ratio of the unit cross sections for the transitions to the ground state and the state at E$_{x}$=1.05 MeV ($\frac{\hat{\sigma}_{g.s.}}{\hat{\sigma}_{1.05}}$). If this ratio equals unity, the proportionality between B(GT) and cross section at zero degrees is perfectly maintained and a deviation from unity signifies proportionality breaking between the two transitions. In addition, the ratio of the distorted-wave calculations with and without the tensor-$\tau$ component of the effective interaction  was calculated, so that it can be determined to what extent the breaking of the proportionality stems from this source.

\begin{table*}
\caption{\label{tab:dw} Results of a theoretical study of the discrepancy between the ratio of the extracted B(GT)s for the transitions to the ground state and first-excited state at 1.05 MeV in $^{58}$Co for the $^{58}$Ni(p,n) and $^{58}$Ni($^{3}$He,t) reactions. The OBTDs that are used in the calculations are from Ref. \cite{HON06}. `PW' and 'DW' refer to the cross section calculated in the plane-wave and distorted-wave approximation, respectively. `DW$_{T\tau=0}$ refers to cross sections from the distorted-wave calculation in which the tensor-$\tau$ amplitudes in the Love-Franey interaction have been set to 0. For further details, see text.}
\begin{ruledtabular}
\begin{tabular}{lcccccccccccc}
&  \multicolumn{6}{c}{GXPF1\footnotemark[1]} & \multicolumn{6}{c}{KB3G\footnotemark[1]} \\
\cline{2-7} \cline{8-13} \\
&  \multicolumn{3}{c}{$^{58}$Ni(p,n)$^{58}$Cu} & \multicolumn{3}{c}{$^{58}$Ni($^{3}$He,t)$^{58}$Cu} &  \multicolumn{3}{c}{$^{58}$Ni(p,n)$^{58}$Cu} & \multicolumn{3}{c}{$^{58}$Ni($^{3}$He,t)$^{58}$Cu}\\
\cline{2-4} \cline{5-7} \cline{8-10} \cline{11-13} \\
& g.s. & 1.05 MeV & $\frac{\hat{\sigma}_{g.s.}}{\hat{\sigma}_{1.05}}$  & g.s. & 1.05 MeV & $\frac{\hat{\sigma}_{g.s.}}{\hat{\sigma}_{1.05}}$ & g.s. & 1.05 MeV & $\frac{\hat{\sigma}_{g.s.}}{\hat{\sigma}_{1.05}}$ & g.s. & 1.05 MeV & $\frac{\hat{\sigma}_{g.s.}}{\hat{\sigma}_{1.05}}$ \\
\hline
B(GT)$_{th}$                   & 0.275 & 0.318 &     & 0.275 & 0.318  &     & 0.496 & 0.768 &     & 0.496 & 0.768 &    \\
\\
$\frac{d\sigma}{d\Omega}(q=0)$ & & & & & & & & & & & & \\
(mb/sr) & & & & & & & & & & & & \\ 
\cline{1-1}
PW                             & 4.52 & 5.28  & 1.00 & 36.9 & 42.9 & 1.00 & 8.19 & 12.8 & 1.00 & 66.5 & 103. & 1.00 \\ 
DW                             & 1.61 & 1.58  & 1.19 & 4.01 & 3.24 & 1.44 & 2.77 & 3.73 & 1.15 & 6.73 & 7.80 & 1.34 \\
DW$_{T\tau=0}$                 & 1.21 & 1.49  & 0.94 & 3.33 & 3.62 & 1.07 & 2.25 & 3.63 & 0.96 & 5.78 & 8.60 & 1.04 \\
\cline{1-1}
DW/DW$_{T\tau=0}$              & 1.33 & 1.06  &      & 1.20 & 0.89 &      & 1.23 & 1.03 &      & 1.16 & 0.91 &      \\
\end{tabular}
\end{ruledtabular}
\footnotetext[1]{In the calculations shown in this table, the strengths and cross sections have not been adjusted for the Gamow-Teller quenching factor of $0.74^{2}$ \cite{MAR96}}. 
\end{table*}

The results of the calculations are summarized in Table \ref{tab:dw}. In the absence of distortions (plane-wave calculation; PW) the proportionality is perfect because all components with $\Delta L=2$, $\Delta S=1$ are zero at q=0. In the full distorted-wave (DW) calculation with the GXPF1 interaction, the ratios of unit cross sections ($\frac{\hat{\sigma}_{g.s.}}{\hat{\sigma}_{1.05}}$) are equal to 1.19 and 1.44 for the (p,n) and ($^{3}$He,t) reactions, respectively. Those numbers are similar when using KB3G: 1.15 and 1.34. In brief, significant proportionality breaking effects are seen in both reactions, and the effect on the ratio of the cross sections for the excitation of the first two $1^{+}$ states is about twice as large for the ($^{3}$He,t) reaction as for (p,n). Therefore, according to the calculations with GXPF1:
\begin{equation}
\label{eq:ratiogx}
\left[\frac{[\frac{\hat{\sigma}_{g.s.}}{\hat{\sigma}_{1.05}}]_{\text{(p,n)}}}{
[\frac{\hat{\sigma}_{g.s.}}{\hat{\sigma}_{1.05}}]_{^{3}\text{(He,t)}}}
\right]_{\text{GXPF1}}=\frac{1.19}{1.44}=0.826
\end{equation}
and with KB3G:
\begin{equation}
\label{eq:ratiokb}
\left[\frac{[\frac{\hat{\sigma}_{g.s.}}{\hat{\sigma}_{1.05}}]_{\text{(p,n)}}}{
[\frac{\hat{\sigma}_{g.s.}}{\hat{\sigma}_{1.05}}]_{^{3}\text{(He,t)}}}
\right]_{\text{KB3G}}=\frac{1.15}{1.34}=0.858
\end{equation}
The ratio [$\frac{\hat{\sigma}_{g.s.}}{\hat{\sigma}_{1.05}}]_{\text{(p,n)}}/[\frac{\hat{\sigma}_{g.s.}}{\hat{\sigma}_{1.05}}]_{^{3}\text{(He,t)}}$
calculated from the experimental results in Table \ref{tab:cu} equals 0.33/0.58=0.57 if the (p,n) data taken at 120 MeV are used and 0.41/0.58=0.7 if the (p,n) data taken at 160 MeV/nucleon are used. In short, the discrepancy between the (p,n) and ($^{3}$He,t) data are qualitatively explained. It is hard to reach stronger conclusions about the quantitative agreement, owing to the $~15$\% error bars in the (p,n) data. New $^{58}$Ni(p,n) experiments at $E_{p}=200$ MeV and 300 MeV have recently been performed \cite{SAS05}, but not yet fully analyzed. The results could shed further light on the analysis. 

From Table \ref{tab:dw} it can be seen that the changes in the unit cross section solely due to the inclusion of tensor-$\tau$ components in the interaction are predicted to be slightly stronger for (p,n) than for ($^{3}$He,t) since the ratio DW/DW$_{T\tau=0}$ is higher for the transitions with the former probe. However, for the (p,n) reaction, these contributions counteract the proportionality breaking due to other causes (such as exchange and $\Delta L=2$, $\Delta S=1$ contributions mediated through the $\sigma\tau$ component of the interaction) on the level of 4-6\%, whereas for the ($^{3}$He,t) reaction they reinforce such effects. In addition, the sign of the interference is the same for both transitions in the case of the (p,n) reaction and opposite in the case of the ($^{3}$He,t) reaction. Therefore, the ratio of unit cross sections is more strongly affected in the case of the ($^{3}$He,t) reaction. 

Besides the calculations in Table \ref{tab:dw}, we also checked the effect of removing all contributions from the $\nu p_{3/2}$-$\pi f_{5/2}$ and $\nu f_{5/2}$-$\pi p_{3/2}$ components in the cross section calculations. The value of ($\frac{\hat{\sigma}_{g.s.}}{\hat{\sigma}_{1.05}}=1$) became 0.91 for the (p,n) reaction and 0.95 for the ($^{3}$He,t) reaction, with very minor differences between the results with KB3G or GXPF1. This confirms that these particle-hole components are indeed the leading cause for the discrepancies between the two reactions and the breaking of the proportionality. 

A systematic error in the absolute scale of the Gamow-Teller strengths extracted from $^{58}$Ni($^{3}$He,t) directly translates into a systematic error in the absolute scale of the strengths extracted from $^{58}$Ni(t,$^{3}$He). Since, according to the calculations, the cross section of the transition to the ground state in $^{58}$Cu is increased due to the interference between $\Delta L=2$, $\Delta S=1$ and $\Delta L=0$, $\Delta S=1$ amplitudes by about 20-25\% (DW/DW$_{T\tau=0}$=1.2 and an additional 5\% is included to account for other proportionality breaking effects), the B(GT) for the other states in the $^{58}$Cu spectrum (including the T$_{>}$ used for calibrating the B(GT) in $^{58}$Co) are underestimated by the same percentage. The same effects occur for the results from the (d,$^{2}$He) experiment for which a similar calibration procedure was used. 

\section{Electron-capture rates}
To understand how the differences between the various measured and calculated Gamow-Teller strength distributions affect the electron capture rates, we calculated these rates for various stages during stellar evolution. The method to calculate electron-capture rates is described in detail in Refs. \cite{fuller80,fuller82a,fuller82b,fuller85} and implemented in a new code \cite{GUP06}. Calculations were performed in a grid spanning $\rho Y_{e}$ values from $10^{1}$ gcm$^{-3}$ to $10^{14}$ gcm$^{-3}$ and T values from $0.01\times 10^9$ K to $100\times 10^9$ K. Here, we present the results for two different $\rho Y_{e}$ regimes of relevance in the late stages of evolution of massive stars (11-40 solar masses) \cite{heger01a}: $\rho Y_{e}=10^{7}$ gcm$^{-3}$, which corresponds to conditions during Silicon burning and Silicon depletion ($Y_{e}\approx 0.47$; this regime was also investigated in Refs. \cite{hagemann04,hagemann05}) and $\rho Y_{e}=10^{9}$ gcm$^{-3}$, which corresponds to conditions during the pre-supernova stage ($Y_{e}\approx 0.44$; $\sim 0.5$ s before core bounce). In the former case, the temperature is $T\sim 4\times 10^{9}$K and in the latter case, $T\sim 8\times 10^{9}$K.
In the present calculations, we only consider transitions from the parent ground state to daughter states described by Gamow-Teller strength distributions and ignore transitions from thermally populated parent states. Especially at the higher temperatures, this leads to an underestimation of the electron-capture rates \cite{langanke00}. However, here we are mostly interested in relative deviations in the rates due to the differences in strength distributions.

\begin{figure}
\begin{center}
\includegraphics[scale=1.0]{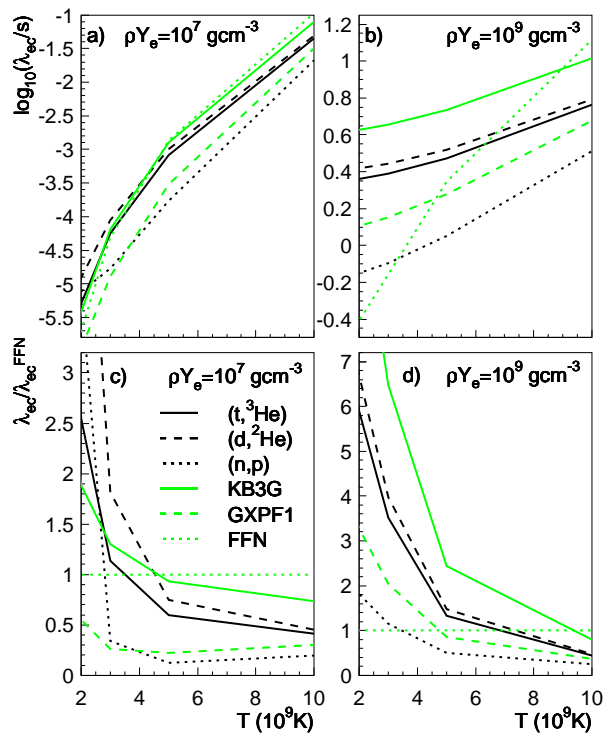}
\caption{a) Electron-capture rates on $^{58}$Ni in the stellar environment at $\rho Y_{e}=10^{7}$ gcm$^{-3}$ using various theoretical (green lines) and experimental (black lines) B(GT) distributions in $^{58}$Co, as labeled in c). b) Same a a) but calculated at $\rho Y_{e}=10^{9}$ gcm$^{-3}$. c) The ratio of electron capture rates in a) to the FFN rates. d) The ratio of electron capture rates in b) to the FFN rates. Note the difference in scales on the Y-axis for the plots. For details, see text. }
\label{rate}
\end{center}
\end{figure}

In addition to the original results by Fuller, Fowler and Newman (FFN) \cite{fuller80,fuller82a,fuller82b,fuller85,ffn}, electron-capture rates were calculated for five Gamow-Teller strength distributions in $^{58}$Co: i) The theoretical prediction employing the GXPF1 interaction \cite{hagemann05}, ii) the theoretical prediction employing the KB3G interaction \cite{hagemann04,hagemann05,MP}, iii) the distribution extracted from the $^{58}$Ni(t,$^{3}$He) experiment, iv) the distribution extracted from the $^{58}$Ni(d,$^{2}$He) experiment \cite{hagemann04,hagemann05} and v) the distribution extracted from the $^{58}$Ni(n,p) experiment \cite{elkateb94}. For cases i) and ii) the $1^{+}$ states in $^{58}$Co were positioned at exactly their calculated values. For case iii), the strength extracted for each 250-keV wide bin was placed at the center of that bin. For case iv), the strength was distributed according to the values extracted state-by-state in Table 1 of Ref. \cite{hagemann04} below E$_{x}$($^{58}$Co)=4. MeV. Above that energy, values extracted per 1-MeV bin (Table 2 of Ref. \cite{hagemann04}) were equally distributed over the respective bins, and placed at the center of four 250-keV bins. Finally, for case v) the strengths extracted per 1-MeV bin were equally distributed over the respective bins and placed at the center of four 250-keV bins.   

The results of the calculations are plotted in Fig. \ref{rate}. Fig. \ref{rate}a shows the rates calculated at the lower densities, at temperatures T$=2-10\times 10^{9}$K and Fig. \ref{rate}c displays the same calculations, but relative to the FFN results (note that the quenching factor of 0.74$^{2}$ is not included in the FFN values). At this low density, and correspondingly low electron chemical potential ($\phi_e\approx 0.7$ MeV at T$=1.0\times 10^{9}$), the rates rise rapidly with temperature due to the fact that electrons can have energies $\sim k_{B}T$ larger than $\phi_e$  ($k_{B}T=90$ keV (900 keV) at T$=1\times 10^{9}$K ($10 \times 10^{9}$K), with $k_{B}$ the Boltzmann constant) and thus increasingly populate the lowest-lying $1^{+}$ states. This thermal broadening of the electron Fermi surface is characterized by the degeneracy parameter $\phi_e/(k_BT)$ (here defined positive) which is reduced (``lifting'' of degeneracy) from $\approx7.7$ at T$=1\times 10^{9}$K to $\approx-0.35$ at  T$=10\times 10^{9}$K. The temperature dependences of the rates at the low density are, therefore, very sensitive to the precise location of these low-lying states, and when calculated from experimental strength distributions, also to the binning and resolution of the data. The rate calculation using the strength distribution obtained with the GXPF1 interaction result in much lower rates than the one employing KB3G at this density, owing to the near absence of strength in the capture window ($\approx (\phi_e+m_ec^2-w)$ where $w$ is the ground-state-to-ground-state capture threshold and  $m_ec^2$ the electron rest mass). Except for the lowest temperatures, the rates calculated with the strength distribution from the (n,p) experiments are relatively low compared to those from (d,$^2$He) and (t,$^3$He) experiments.

At the higher density,  $\phi_e\approx 4.7$ MeV. As a result, a larger fraction of the strength distribution can be accessed compared to the case at lower densities. The degeneracy $\phi_e/(k_B T)$ reduces from $\approx$52 at T$=1\times 10^{9}$K to $\approx$4.66 at T$=10\times 10^{9}$K, but the Fermi-Dirac distribution still resembles a sharp energy filter that accesses daughter states almost exclusively in the capture window. The dependence of rate on temperature is thus relatively weak, as shown in Fig. \ref{rate}b. Rate estimates using strength distributions that have relatively little strength located within the capture window (i.e. the one extracted from the (n,p) experiment and the theoretical calculation using the GXPF1 interaction) will result in lower rates than those exhibiting more Gamow-Teller strength at low excitation energies (i.e. the (t,$^{3}$He) and (d,$^{2}$He) experimental results and the theoretical calculation with the KB3G interaction), especially since the phase space available for capture depends strongly (power of $\sim 5$) on the total electron energy. The FFN rates have a significantly steeper dependence on temperature.

\section{Conclusions}
\label{sec:conclusions}
The Gamow-Teller strength distribution in $^{58}$Co has been extracted from the $^{58}$Ni(t,$^{3}$He) reaction at 115 MeV/nucleon. Although the statistical errors are relatively large, the results are important to cross check existing results from $^{58}$Ni(d,$^{2}$He) and $^{58}$Ni(n,p) experiments, which were inconsistent. Our data are consistent with those from the $^{58}$Ni(d,$^{2}$He) reaction and deviate from the $^{58}$Ni(n,p) data.

Comparisons between the experimentally extracted strength distributions and shell-model calculations using the GXPF1 and KB3G interactions were made. The predictions with the KB3G interaction describe the strength distributions at excitation energies below 4 MeV in $^{58}$Co well, but the calculations with the GXPF1 interaction better reproduce the strength distribution at higher excitation energies.    

Systematic uncertainties in the calibration of the absolute Gamow-Teller strength scale were investigated. Such uncertainties are due to potentially large interference effects between $\Delta L=2$, $\Delta S=1$ and $\Delta L=0$, $\Delta S=1$ components for the transition $^{58}$Ni($^{3}$He,t)$^{58}$Cu(g.s.) which is used in the calibration. A correction for this effect would increase the Gamow-Teller strengths extracted  from $^{58}$Ni(t,$^{3}$He) (and $^{58}$Ni(d,$^{2}$He), since the same procedure for the strength calibration was used) by about 25\%.

Finally, the differences between the various experimental and theoretical strength distributions were investigated in terms of electron-capture rates in the stellar environment. At low densities and corresponding narrow electron-capture window, the rates are very sensitive to the details of the strength distribution at the lowest excitation energies and, therefore, the rates calculated with the strength distributions predicted using the KB3G interaction are closest to those predicted using experimentally determined strength distributions from $^{58}$Ni(t,$^{3}$He) and $^{58}$Ni(d,$^{2}$He). At higher densities, the electron-capture window encompasses a large portion of the Gamow-Teller strength distribution and the rates depend much more on the mean location and the width of the Gamow-Teller strength distribution and rates calculated using the strength distributions from shell-model calculations with the GXPF1 or KB3G interactions do about equally well in reproducing the rates calculated with experimental strength distributions from the $^{58}$Ni(t,$^{3}$He) and $^{58}$Ni(d,$^{2}$He) data.

\begin{acknowledgments}
We thank the cyclotron staff at NSCL for their support during the experiment described in this paper.
This work was supported by the US NSF (PHY02-16783 (JINA), PHY-0110253), the Ministry of Education, Science, Sports and Culture of Japan, the Stichting voor Fundamenteel Onderzoek der Materie (FOM), the Netherlands and by the Office of the Vice President for Research, University of Michigan. One of us, R.Z., wishes to thank G. Mart\'{i}nez-Pinedo and M. Honma for providing results from their shell-model calculations and, together with K. Langanke and B.A. Brown, for discussions about the contents of the paper.
\end{acknowledgments}   

\bibliography{prcni}

\end{document}